# scBIT: Integrating Single-cell Transcriptomic Data into fMRI-based Prediction for Alzheimer's Disease Diagnosis

Yu-An Huang[1,2, #,*], Yao Hu[3, #], Yue-Chao Li[1], Xiyue Cao[1], Xinyuan Li[1], Kay Chen Tan[3], Zhu-Hong You[1,*], Zhi-An Huang[4,*]

[1] School of Computer Science, Northwestern Polytechnical University, Shaanxi 710129, China

[2] Research & Development Institute of Northwestern Polytechnical University in Shenzhen, Shenzhen, 518063, China

[3] Department of Data Science and Artificial Intelligence. The Hong Kong Polytechnic University, Hong Kong, SAR, China

[4] Research Office, City University of Hong Kong (Dongguan), Dongguan 523000, China

#equal contribution

*corresponding authors

## Abstract

Functional MRI (fMRI) and single-cell transcriptomics are pivotal in Alzheimer's disease (AD) research, each providing unique insights into neural function and molecular mechanisms. However, integrating these complementary modalities remains largely unexplored. Here, we introduce scBIT, a novel method for enhancing AD prediction by combining fMRI with single-nucleus RNA (snRNA). scBIT leverages snRNA as an auxiliary modality, significantly improving fMRI-based prediction models and providing comprehensive interpretability. It employs a sampling strategy to segment snRNA data into cell-type-specific gene networks and utilizes a self-explainable graph neural network to extract critical subgraphs. Additionally, we use demographic and genetic similarities to pair snRNA and fMRI data across individuals, enabling robust cross-modal learning. Extensive experiments validate scBIT's effectiveness in revealing intricate brain region-gene associations and enhancing diagnostic prediction accuracy. By advancing brain imaging transcriptomics to the single-cell level, scBIT sheds new light on biomarker discovery in AD research. Experimental results show that incorporating snRNA data into the scBIT model significantly boosts accuracy, improving binary classification by 3.39% and five-class classification by 26.59%. The codes were implemented in Python and have been released on GitHub (https://github.com/77YQ77/scBIT) and Zenodo (https://zenodo.org/records/11599030) with detailed instructions.

# 1. Introduction

Alzheimer's disease, a progressive neurological disorder impacting cognitive function and daily life, affects an estimated 50 million people worldwide and is diagnosed using genetic sequencing and medical imaging techniques [1]. Genetic sequencing provides detailed information about an individual's genome at a microscopic level, including specific genetic variations and genetic risk-related data [2, 3]. Meanwhile, medical imaging techniques offer macroscopic insights, providing detailed images of brain structure, metabolic activity, and identifiable areas of abnormality, aiding in the identification of lesions, atrophy, and other visible brain changes [4, 5]. Both imaging and genetic information are vital markers for diagnosing AD [6, 7]. Their combined use has led to the emergence of studies such as imaging genomics and imaging transcriptomics, significantly contributing to a more comprehensive understanding of the disease's underlying mechanisms at both genetic and neuroimaging levels [8, 9].

Imaging genomics and transcriptomics represent advanced interdisciplinary methodologies that integrate genetic and neuroimaging datasets, aiming to elucidate the correlations between genomic variations, transcriptomic profiles, and neuroimaging phenotypes [10, 11]. The foundational hypothesis posits that distinct gene expression signatures are associated with specific neuroimaging characteristics, thereby enhancing the pathophysiological understanding of neuropsychiatric disorders and interindividual variability. This goes beyond what can be derived from imaging alone, which typically reveals structural or functional alterations without direct molecular context. In comparison with gene sequencing-based diagnostic methods, these approaches confer a notable advantage over traditional gene sequencing-based diagnostic methods by potentially obviating the necessity for invasive biopsy procedures, thereby reducing the attendant morbidity and mortality risks [12, 13]. Furthermore, they facilitate a comprehensive spatiotemporal analysis of tumor heterogeneity, an endeavor that is unattainable via conventional serial biopsy methods.

Genome-wide association studies (GWASs) now involve over a million participants and analyze millions of genetic variants throughout the genome, identifying dozens to thousands of genetic variants statistically linked to various diseases including AD. However, owing to the 'missing heritability' phenomenon, which is partly attributed to pleiotropy, polygenicity, and coarse phenotype resolution, these GWAS variants often do not correspond to functional genetic variants [14]. With the use of imaging data, imaging transcriptomics is able to delineate AD imaging-derived phenotypes, presenting new opportunities for identifying functional variants associated with more precise and fine-grained phenotypes. While imaging genomics delves into the genetic variants related to imaging data, the recent advent of imaging transcriptomics explores gene expression patterns across the brain and relate these patterns to various structural and functional properties as quantified by neuroimaging techniques, providing a powerful tool for understanding the molecular mechanisms underlying AD from an alternative perspective [15, 16].

Using atlases of whole-brain gene expression data, research in this field aims to uncover spatial gene expression patterns that correlate with brain structure and function, thereby providing insights into the molecular mechanisms underlying brain organization, and facilitating the diagnosis of neurological disorders such as AD. However, as emphasized by Mandal et al. [17], several significant limitations within current imaging transcriptomics studies are impeding their application. Firstly, while gene set enrichment is a routine practice in current imaging transcriptomic studies, its analysis results lack interpretation partly because the molecular pathways they reveal are derived from post hoc enrichment analyses rather than being directly computed from the expression data. Secondly, current analyses in imaging transcriptomics face challenges in fully integrating cell type information, which is pivotal as it typically represents the primary source of variation across bulk brain transcriptomic datasets. Finally, existing datasets encounter the challenge of limited sample sizes, exemplified by the Allen Human Brain Atlas (AHBA) [18], the most comprehensive anatomical expression atlas, which is derived from data from merely 6 individuals, predominantly from the left hemisphere. As the accumulation of data from single-cell molecular profiling technologies related to AD increases [19], it offers a hopeful prospect by providing crucial supplementary information for imaging transcriptomics, to some extent alleviating issues related to data sources. It introduces a new level of single-cell expression data, providing a foundational basis for comprehensive molecular pathway characterization, including gene regulatory networks. Additionally, it inherently preserves differential expression across cell types and expand the sample size of patient data. The integration of snRNA sequencing and neuroimaging data effectively addresses the current challenges encountered by imaging transcriptomics from a data-sourcing perspective, while simultaneously presenting significant computational challenges that warrant careful consideration [17]. The foremost challenges lie in how to match the two modalities of data from different patients and how to coarsen the expression information from individual genes at the single-cell level to a meaningful feature representation at an appropriate granularity, and subsequently align it with neuroimaging features.

In this article, to address these challenges, we introduce the scBIT(acronym for single-cell Brain Imaging Transcriptomics) model, which innovatively combines single-cell transcriptomics data with rs-fMRI data for the diagnosis of AD. Based on a cell-type-scale gene interaction subgraph representation method, the scBIT model can learn the attention between brain region and gene subgraphs through contrastive learning and subsequently predict the probability of AD using any fMRI data without any prior hypothesis. The proposed approach unveils a pioneering perspective for exploring brain region and gene expression patterns on a single-cell or single-nucleus scale, while introducing an innovative framework that harnesses the power of snRNA-fMRI cross-modal data for the diagnosis of AD. The main contributions of this work are summarized as follows:

(1) We introduce scBIT, the first computational model to integrate single-cell RNA data with MRI data for AD prediction, providing a new approach for computational models in neurodegenerative disease research.

(2) We employ cross-modal contrastive learning for unpaired individuals, which not only

improves diagnostic precision but also deepens the interpretability of pathogenic gene networks and their connections with brain regions and cell types.

(3) We propose a new feature extraction method for snRNA-seq data, which is based on a bagging strategy to obtain coarse-grained features that preserve cell type and gene relationship information in single-cell data for attention computation with brain region features.

(4) We implement similarity computations for patient data across snRNA-seq and fMRI data types, enabling the identification of relevant cases across diverse datasets.

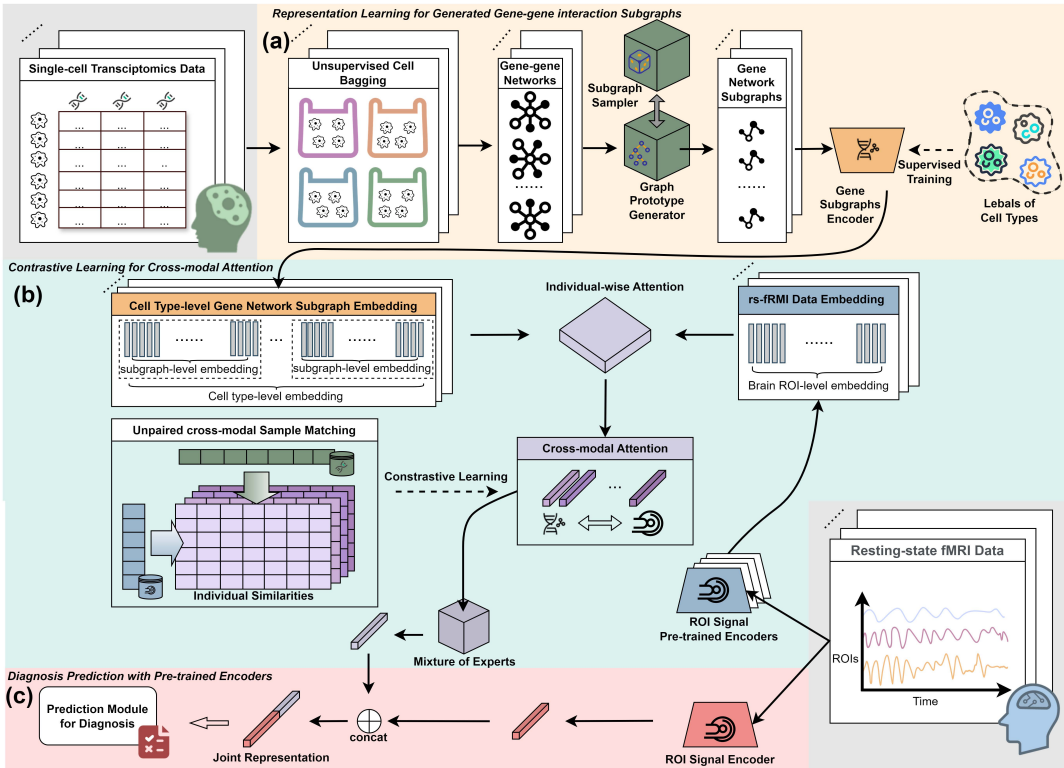

**Fig. 1.** Overview of the scBIT framework.

## 2. Related works

Since there are currently no studies combining single-cell data with MRI data for predicting AD, this section focuses on methods that exclusively utilize MRI data for AD classification or prediction. Recent research in AD prediction using neuroimaging data incorporates advanced methodologies that include network techniques and machine learning algorithms. Mao et al. [20] leveraged rs-fMRI to extract ALFF and ReHo metrics, studying physiological changes across brain regions. Tripoliti et al. [21] employed fMRI data with random forest algorithms to develop a robust disease classification approach. Hojjati et al. [22] proposed a method that combines structural and functional MRI to identify potential MCI progressors, demonstrating the utility of multimodal imaging.

Further advancements have been made through the use of innovative neural network models. Zhang et al. [23] introduced a multi-scale time series kernel learning model specifically designed for diagnosing brain diseases from fMRI data. Bi et al. [24] implemented stochastic neural network clustering to analyze fMRI data. Deep learning approaches were further explored by Gupta and Odusami [25, 26], who utilized 2D-CNN models and ResNet18 frameworks, respectively, to process MRI slices. Abrol et al. [27] demonstrated the effectiveness of combined sMRI and fMRI models in boosting predictive accuracy. Additionally, dynamic frequency functional networks were examined by Khatri et al. [28].

## 3.  Methodology

### 3.1 Problem formulation

In this article, the prediction task of AD diagnosis using snRNA and fMRI multimodal data is designed with snRNA as the helper modality. Through contrastive learning, correlations across different samples from the datasets are constructed, and these correlations are then integrated with fMRI data, which serves as the primary modality, to predict clinical state from imaging data. Specifically, for a set of single-cell transcriptomic data from $N_t$ individuals, denoted as $\{T_i\}_{i=1}^{N_t}$, each $T_i$ includes a gene expression matrix of dimensions determined by the number of nuclei and the number of genes, with each nucleus labeled by its cell type. Accompanying each individual's data are demographic and clinical labels such as age, gender, and disease state. Given an fMRI dataset consisting of N individuals, $\{X_i\}_{i=1}^N$, the first objective is to train separate encoders for two modalities, represented as $T'_i = E_{sn}(T_i)$ for snRNA data and $X'_i = E_{rm}(X_j)$ for fMRI data. An attention mechanism is then employed to learn attention scores $a_j$ that quantify the relevance of the entire snRNA dataset to each fMRI image $X_j$. Subsequently, a classification function $f$ is constructed to use the fMRI data and the derived attention scores to compute the disease probability for each image: $s_j = f(X'_j, a_j)$. This approach aims to harness the complementary information from both data modalities through a sophisticated learning scheme, emphasizing cross-modal correlations to enhance the accuracy of AD prediction.

## 3.2 Overview of scBIT

scBIT is an end-to-end and hypotheses-free framework designed to predict the probability of AD using any fMRI data, supplemented with snRNA data as an auxiliary modality. It facilitates interpretative analysis by delineating the intricate relationships between macroscopic brain regions and microscopic genetic networks. There are three steps in scBIT framework (**Fig. 1** and Methods). (i) Each snRNA dataset is initially processed through a bagging strategy to transform into 'cell bags', where each bag contains cellular gene expression data to form individual gene networks. A self-explainable graph neural network is then employed to predict cell types for each gene network and extract key gene subgraph representations that significantly determine cell type classification. Each cell type is represented by an equal number of embeddings. Through this approach, the snRNA data matrix for each individual can be converted into subgraph-level embeddings. (ii) Given that snRNA datasets and fMRI datasets are collected from different individuals, associations between unpaired individuals across these datasets are constructed using a variety of matching strategies. These strategies build on similarities based on age, gender, genetics, and clinical state to enable cross-modal contrastive learning. For each type of similarity, a specific encoder is trained on the fMRI data to derive embeddings at the brain region ROI level. This training facilitates the computation of four distinct types of attention scores between the fMRI data and each snRNA dataset. A mix-of-experts [29] (MoE) model is then utilized to integrate these four attention scores into a unified score, representing the relevance of the fMRI data to each gene expression matrix in the snRNA dataset. (iii) An encoder is trained on the fMRI training dataset to generate embeddings that encapsulate the functional attributes of brain regions. These embeddings are integrated with cross-modal individual-level attention scores, previously computed in step 2, to form a joint representation. This representation is then analyzed using a classifier to derive predictive outcomes for AD diagnosis.

## 3.3 scBIT architecture

The overall model architecture is depicted in **Fig. 1**, which contains three stages: (1) representation learning for generated gene-gene interaction subgraph; (2) contrastive learning for cross-modal attention; (3) diagnosis prediction with pre-trained encoders. The specific implementation details of this model in this work are presented in **Fig. 2**.

**Stage 1: Representation learning for generated gene-gene interaction subgraph.** scBIT utilizes a bagging strategy to construct sets of gene subgraphs from snRNA expression matrices, and it employs a self-explainable graph neural network framework (**Fig. 2(a)**) for model training, using cell types in snRNA as labels. This method takes the interpretable sets of subgraphs as representations for each cell type in snRNA to achieve coarse-grained feature representation. Given a set of single-cell transcriptomic data from $N_t$ individuals, $\{T_i\}_{i=1}^{N_t}$, this stage transforms each $T_i$ into a collection of cell type level gene network subgraph embeddings, $\{p_j\}_{j=1}^{N_c \times N_s}$, where $N_c$ is the number of cell types (which equals 15 in this work) and $N_s$ is the predefined number of gene subgraphs used to represent each cell type (set to 6 in this work). Here, each embedding $p_j$ is interpretably mapped to a gene subgraph, with which it corresponds one-to-one.

**Cell bagging for gene-gene network construction.** To characterize the distinct gene interaction patterns inherent to each cell type within snRNA, scBIT converts snRNA expression data into a collection of gene networks. Given an snRNA matrix $T_i$, scBIT randomly selects a fixed number of cell nuclei (set to 20 in this work) from the same cell type to place into the same bag. Within this bag, the PCC value of each gene-gene pair across the 20 cell nuclei is calculated, and the top 20% pairs with the highest PCC value are retained to construct a gene-gene network. This process is repeated until all cells of the same type have been processed. Consequently, the matrix $T_i$ (comprising $N_n$ cell nuclei) is converted into a set of gene networks $\{g_j\}_{j=1}^{m}$ ($m = \left\lfloor \frac{N_n}{20} \right\rfloor$), wherein each node within $g_i$ has node features, represented by the expression values of the respective gene across the 20 cell nuclei within the $j$-th bag. Since the cell nuclei in the same bag are sampled from the same cell type, the label of network they constitute also corresponds to the shared cell type.

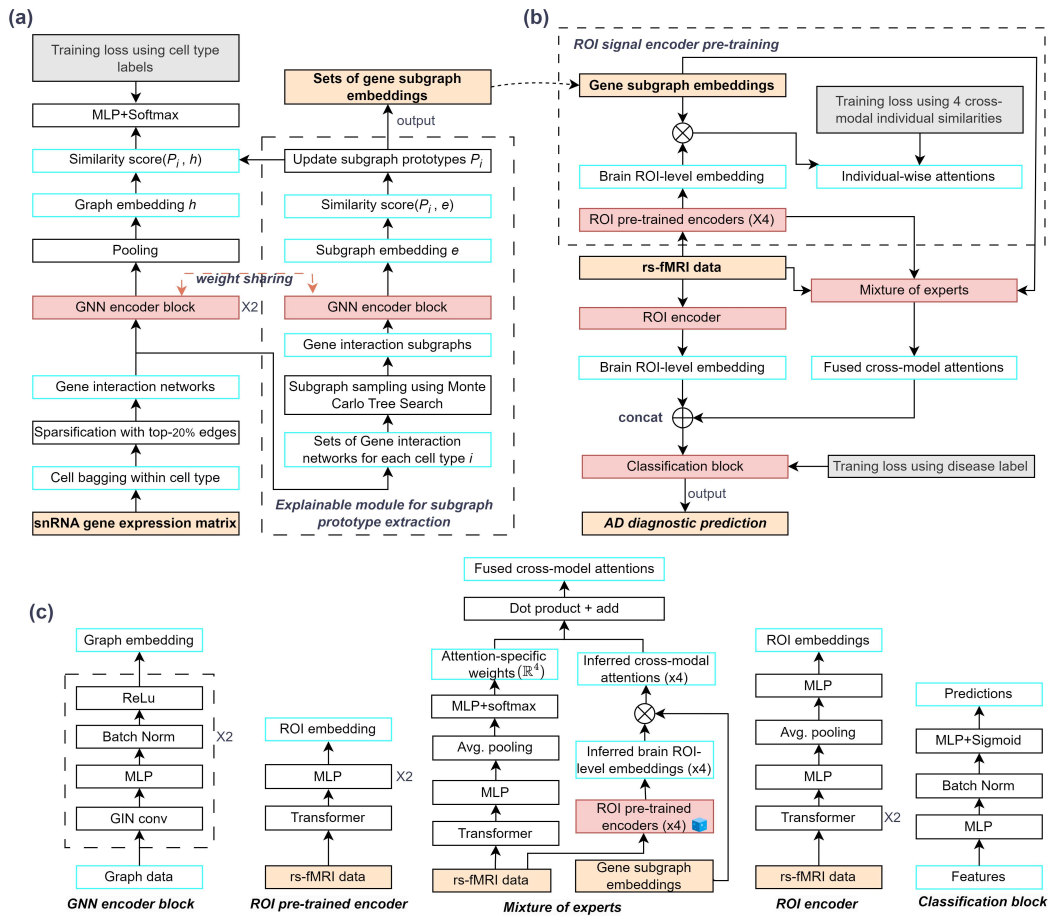

**Fig. 2.** Schematic illustration of the scBIT model architecture. (a) Schematic of gene subgraph embedding learning from snRNA data using an interpretable graph neural network. (b) Schematic of the AD diagnostic prediction model based on contrastive learning. (c) Detailed implementation of components within the scBIT model.

**Self-explainable GNN for gene subgraph extraction.** As depicted in **Fig. 2(a)**, each gene interaction book is introduced into a self-explainable Graph Neural Network (GNN) as input. The network is structured into two main components. The first component uses a GNN encoder to acquire a graph-wise embedding of the entire gene network. This embedding is then integrated with prototypes, generated in the second component of the network using the interpretability strategy of the ProtGNN method [30]. These prototypes aid in computing the similarities between the gene network and all cell types, which are subsequently processed through an MLP (Multilayer Perceptron) coupled with a softmax function to predict the cell type of the gene network. The second component generates six prototypes for each cell type and utilizes Monte Carlo Tree Search (MCTS) [31] to identify the subgraphs that correspond to these prototypes. We let $\mathcal{G} = \{G_i\}_{i=1}^{15}$ be a collection of $m$ gene networks from an snRNA-seq matrix, where each $G_i$ consists of $\{g_q\}_{q=1}^{m_i}$, a set of gene networks specific to its cell type. The equations for the architecture of the first component are

$$h_q = \text{AvgPool}(\text{GNN}_{\text{enc}}(g_q)) \tag{1}$$

$$sim(p_i, h_q) = \left\| h_q - p_i \right\|_2 \tag{2}$$

$$d_q = \left[ sim(p_i, h_q) \middle| k = 1 \dots 90 \right] \tag{3}$$

$$\widetilde{y_q} = \text{softmax}(\text{MLP}(d_q)) \tag{4}$$

$$loss_{cls} = \frac{1}{m} \sum_{q=1}^{m} \min_{i : p_i \in P_{y_i}} \left\| h_q - p_i \right\|_2^2 \tag{5}$$

$$loss_{sep} = -\frac{1}{m} \sum_{q=1}^{m} \min_{i : p_i \notin P_{y_i}} \left\| h_q - p_i \right\|_2^2 \tag{6}$$

$$LOSS = \frac{1}{m} \sum_{q=1}^{m} \text{CrsEnt}(\widetilde{y_q}, y_q) + \lambda_1 \, loss_{cls} + \lambda_2 loss_{sep} \tag{7}$$

$\text{GNN}_{\text{enc}}$ is a graph-level encoder shared by the networks in both components, with detailed specifications

provided in **Fig. 2(c)**; $p_i$ is derived from a set of graph prototypes $P$ maintained by the second component, comprising 90 prototypes distributed among 15 cell types, with six prototypes per type; $loss_{cls}$ and $loss_{sep}$ respectively ensure that each gene network embedding is close to at least one prototype of its own cell type and distant from prototypes of other cell types; $\lambda_1$ and $\lambda_2$ were set to 0.1 and 0.05 in this work.

To generate and maintain the set of prototypes $P = \{P^1, \ldots, P^i \ldots, P^{15}\}$ where each $P^i = \{p_{6 \times i-5}, \ldots p_{6 \times i}\}$, the second component utilizes the MCTS algorithm to optimize the search for subgraphs and to project these prototypes onto the most representative subgraphs that correspond to their respective cell types (**Fig. 2(a)**). To enhance training time efficiency, such subgraph search and prototype projection are performed at fixed intervals during the training process. $P$ is initially randomized. The equations for the architecture of the second component are

$$G_i' = \text{MCTS}(G_i) \tag{8}$$

$$e_{g'} = \text{GNN}_{\text{enc}}(g_i') \quad \text{where} \ g_i' \in G_i' \tag{9}$$

$$sim(p_i, e_{g'}) = \log \left( \frac{\left\| p_i - e_{g'} \right\|_2^2 + 1}{\left\| p_i - e_{g'} \right\|_2^2 + \epsilon} \right) \quad \text{where} \ p_i \in P^i \tag{10}$$

$$p_i \leftarrow \arg\min_{e_{g'}} sim(p_i, e_{g'}) \tag{11}$$

MCTS guides the iterative search for the nearest subgraphs through multiple iterations, with each iteration involving a forward phase for path selection and a backward phase for updating statistics, initially favoring the exploration of less-visited nodes to optimize pruning actions, and ultimately selecting the subgraph with the highest similarity to the prototype as the new projected prototype. By using Equation 9, we can conceptually map each prototype to a subgraph, thus enabling interpretative traceability from the prototype to the subgraph. $\epsilon$ was set to 1e-4. In summary, by integrating the two components of this self-explainable GNN model, each snRNA data $T_i$ can be transformed into a set $P$, which consists of 90 prototypes representing 15 different cell types.

**Stage 2: Contrastive learning for cross-modal attention.** scBIT builds different ROI pre-trained encoders to extract brain ROI-level embeddings, and it calculates the cross-modal attentions among unpaired individuals. Then, contrastive learning is employed to train the ROI pre-trained encoders with four different cross-modal individual similarities as labels, which optimizes the similarities between ROI-level and gene subgraph embeddings from unpaired samples to realize cross-modal matching. Given the calculated gene subgraph embedding sets of entire snRNA dataset $\mathcal{P}$ and a set of fMRI data from $N$ individuals, $\{X_i\}_{i=1}^{N}$, this stage first embeds each $X_i$ into brain ROI-level embeddings, $\{\hat{X}_i\}_{i=1}^{N}$, and calculates similarity between $\mathcal{P}$ and $\{\hat{X}_i'\}_{i=1}^{N}$ as the cross-modal attention $a_i$.

**Similarity calculation for individual matching in cross-modal datasets.** For each pair of individuals across datasets, we calculate similarity, $s \in \{\text{age}, \text{sex}, \text{state}, \text{gene}\}$, based on four types of information: age, gender, clinical state, and genetic information. For gender, we assign a value of "1" to males and "-1" to females, then calculate the absolute value of the difference. For age, individuals in the snRNA dataset recorded as "90+" are simply considered as 90, and the difference is then calculated. For clinical state, we assign values of 0, 0.25, 0.5, 0.75 to individuals in the snRNA dataset labeled as "Not AD", "Low", "Intermediate", "High", and values of 0, 0.25, 0.5, 0.75, 1 to those in the fMRI dataset labeled as "CN", "EMCI", "MCI", "LMCI", "AD", respectively, then calculate the difference. For the three types of similarities mentioned above, we performed a normalization process at the end.

In contrast to the gene expression information present in snRNA data, the fMRI dataset solely records individuals' SNP genomic information, necessitating distinct processing approaches for each data type. For the fMRI dataset's SNP data, we used PLINK [32] for preprocessing: filtering out SNPs with high missing rates and those not meeting Hardy-Weinberg equilibrium, removing SNPs with a minor allele frequency under 5%, and pruning linked SNPs. We then annotated SNPs via the NCBI database, summing genes for SNPs linked to multiple genes, and calculated AUCell scores [33] from the KEGG pathway database to assess pathway activities. For the snRNA data, we compute the AUCell scores for each cell across the KEGG pathway dataset and subsequently average these scores. Ultimately, we employ cosine similarity to compare the AUCell scores derived from the two modalities, utilizing this metric as a measure of individual similarity within the genetic information.

**ROI pre-trained encoder for fMRI.** As depicted in **Fig. 2(b)**, to calculate similarity among unpaired individuals within cross-modal datasets, scBIT first need to build pre-trained encoders ($E_{rm}^{pt}$) for fMRI data to extract brain ROI-level embeddings. In this study, the fMRI feature pre-trained encoder is structured into two components, as presented in **Fig. 2(c)**. The first component utilizes two consecutive Transformer encoder layer (TEL) to embed the brain ROI sequence by capturing the long-term dependence via the self-attention mechanism. Then, the output embedding features are fed into one MLP layer in the second component of the encoder network to compile the abstract features for getting the brain ROI-level embeddings. For $s$-th type of similarity, a specific encoder is built on the fMRI data as follows:

$$\hat{X}_i^{s,\prime} = E_{rm}^{pt,s}(X_i) = \text{MLP}\left(\text{TEL}_2^s\left(\text{TEL}_1^s(X_i)\right)\right) \text{ where } s \in \{\text{age}, \text{sex}, \text{state}, \text{gene}\} \tag{12}$$

Subsequently, we devise four different types of attention scores to establish associations among individuals across the datasets based on $\hat{X}_i^{s,\prime}$.

**Contrastive learning for cross-modal matching.** scBIT measures the cosine similarity, $c_{sim}$, between $P_j$ ($P_j \epsilon \mathcal{P}$) and $\hat{X}_i^\prime$ as cross-modal attentions and employs different optimize strategies for matching unpaired snRNA-fMRI samples. Specifically, for the attention scores for discrete cross-modal individual similarities, i.e., the sex and state, scBIT formulates the training of $E_{rm}^{pt,s}$ where $s \in \{\text{sex}, \text{state}\}$ as classification problem by transforming the individual similarity labels into one-hot format.

Then, supervised contrastive learning, $\mathcal{L}_{cl}^s$, is leveraged to train $E_{rm}^{pt,s}$ due to its capability of pulling the leaned embeddings with identical labels together while pushing those from different classes apart, so as to learn more generalizable representations. First, to avoid overfitting, a weak augmentation method with Gaussian noise is performed on $X_i$, resulting in an augmented dataset $\tilde{X}_t$. By leveraging label information, the $\mathcal{L}_{cl}^s$ for $E_{rm}^{pt,s}$ can be formulated as follows:

$$\mathcal{L}_{cl}^s = \sum_{i=1}^{2N} \frac{-1}{|D(j)|} \sum_{j=1}^{2N} \mathbb{1}_{i \neq j} \cdot \mathbb{1}_{\tilde{y}_i = \tilde{y}_j} \cdot \log \frac{\exp\left(\nmid(\hat{\tilde{X}}_i^{s,\prime}) \cdot (\hat{\tilde{X}}_j^{s,\prime})/\tau\right)}{\sum_{k=1}^{2N} \mathbb{1}_{i \neq k} \exp\left(\nmid(\hat{\tilde{X}}_i^{s,\prime}) \cdot (\hat{\tilde{X}}_k^{s,\prime})/\tau\right)} \tag{13}$$

where $\hat{\tilde{X}}_i^{s,\prime}$ denotes the embedding extracted by $E_{rm}^{pt,s}$ from $\tilde{X}_t$ and $\nmid(\cdot)$ denotes normalizing the embeddings into a unit hypersphere. $D(i)$ represents the set of indices of the augmented samples sharing the same label with the $i$-th sample and $\tau \in \mathbb{R}^+$ is the scalar temperature parameter and set to 0.1 in this work. As for the attention scores for continuous similarity values, scBIT take the training of $E_{rm}^{pt,s}$ where $s \in \{\text{age, gene}\}$ as regression tasks by directly using the cross-modal individual similarities as labels. Then, the mean square error between the calculated cross-modal attentions and individual similarities is used to train $E_{rm}^{pt,s}$ where $s \in \{\text{age, gene}\}$.

Subsequently, scBIT calculates cosine similarity between $P$ and $\hat{X}_i^{s,\prime}$ that are inferred by well pretrained $E_{rm}^{pt,s}$ as different types of cross-modal attentions, $a_i^s$, as follows:

$$a_i^s = c_{sim}\left(P, \hat{X}_i^{s,\prime}\right) = c_{sim}(P, E_{rm}^{pt,s}(X_i)) \tag{14}$$

The attention scores can quantify the relevance of the entire snRNA dataset to each fMRI image $X_i$, which are taken as intermodal information to aid the fMRI-based AD diagnosis.

**Stage 3: Diagnosis prediction with pre-trained encoders.** As shown in **Fig. 2(b)** scBIT first builds an ROI encoder, $E_{rm}$, on the fMRI dataset to generate brain ROI-level embeddings, $X'$, that encapsulate the functional attributes of brain regions. To integrate the inter-modal information, scBIT employs a mixture of experts (MoE) to synthesize the cross-modal attentions into a unified score to form a robust representation that captures inter-modal relationships. This method takes the fMRI data and gene subgraph embeddings as inputs and generates attention-specific weights for the inferred cross-modal attentions $a_i^s$ to output the fused cross-modal attentions $A_i$ as follows:

$$A_i = \sum_{s=1}^{4} \text{Softmax}(X_i \cdot W_{\text{MoE}})^s c_{sim}(P, E_{rm}^{pt,s}(X_i)) \tag{15}$$

where $W_{\text{MoE}}$ denotes the learned parameters of MoE model. Then, the fused cross-modal attentions are integrated with the learned fMRI embeddings $X_i'$ as the joint representations. This representation is fed into a classifier, $f$, to derive predictive outcomes for AD diagnosis, $\hat{y}_i$, as follows:

$$\hat{y}_i = \text{Sigmoid}\big(f(\langle X_i', A_i \rangle)\big) \tag{16}$$

where $\langle \cdot \rangle$ denote a concatenation operator.

## 4. Results

### 4.1 Data and materlals

All datasets used in this work are publicly available. In our study, we utilized multiple datasets to provide a comprehensive analysis of AD. Functional MRI data were sourced from the ADNI, which includes imaging data along with demographic information such as age, sex, and disease status of the participants. This data can be accessed through the ADNI Image Collections (https://ida.loni.usc.edu/pages/access/search.jsp). Additionally, we used single-nucleus RNA sequencing data from the Seattle Alzheimer's Disease Brain Cell Atlas, which provides detailed cellular-level information, including demographic and clinical metadata. This data, derived from the MTG and DLPFC, is available at Seattle Alzheimer's Disease Brain Cell Atlas (https://portal.brain-map.org/explore/seattle-alzheimers-disease/seattle-alzheimers-disease-brain-cell-atlas-download?edit&language=en).

To enrich our analysis, we also incorporated genetic information from the ADNI project, which includes data on various genetic markers associated with AD, demographic details such as age and sex, and clinical information. This genetic data can be accessed through the ADNI Genetic Data (https://ida.loni.usc.edu/pages/access/geneticData.jsp?project=ADNI&page=DOWNLOADS&subPage =GENETIC_DATA). These datasets were used to analyze the progression and biological underpinnings of AD, providing a multi-faceted view of the condition through imaging, cellular, and genetic perspectives.

Gene pathway datasets used in this work include KEGG, Reactome, WikiPathways, HumanCyc, PathBank, and Panther. The first three databases can be accessed through the following URLs: KEGG (http://www.kegg.jp/), Reactome (https://www.reactome.org), and WikiPathways (https://www.wikipathways.org). The latter three databases, HumanCyc, PathBank, and Panther, are available via Pathway Commons (https://www.pathwaycommons.org/archives/PC2/v12/). These resources provide comprehensive pathway information that was utilized in our analyses.

## 4.2 scBIT achieves superior performance in Alzheimer's diagnosis with the aid of snRNA data

To evaluate the predictive performance of scBIT, we curated cross-modal datasets from two databases: the Seattle Alzheimer's Disease Cell Atlas (SEA-AD) [34] and the Alzheimer's Disease Neuroimaging Initiative (ADNI) [35, 36], with their statistical data presented in **Fig. 3(e)** and **Fig. 3(f)**. Considering that scBIT's cross-modal contrastive learning framework requires genetic information from fMRI data providers, we exclusively utilized fMRI data from the ADNI dataset for performance testing, despite the availability of other Alzheimer's-related fMRI datasets that do not provide individual genetic information. We employed a ten-fold cross-validation experimental framework to benchmark the prediction performance on the ADNI dataset. This involved using nine folds alternately as the training set, with the remaining fold serving as the test set. To avoid the data leakage problem, we mask the disease state similarity corresponding to the testing set during the training of the pre-trained encoder. We designed two types of prediction tasks: a binary classification task to predict whether an individual has AD and a five-class classification task to predict the severity of the disease. For the binary classification task, we used Accuracy (ACC), Area Under the Curve (AUC), Sensitivity (SEN), and Specificity (SPE) as performance metrics. For the five-class severity assessment, we utilized ACC, F1-Score (F1), Precision (PRE), Recall (REC), and SPE as the evaluation criteria. The assessment was quantified by calculating the mean and standard deviation across the ten folds.

**Fig. 3(a)** illustrates the predictive performance of scBIT on binary and five-class classification tasks demonstrating the impact of incorporating versus omitting snRNA data. The architecture of the encoder and predictor in the scBIT model, when not incorporating snRNA data, aligns with that of the cross-modal version of scBIT. Experimental results indicate that introducing snRNA data into the scBIT model significantly enhances predictive performance, with improvements in the binary classification task of 3.39% in ACC, 5.63% in AUC, 2.74% in SEN, and 9.71% in SPE, and in the five-class classification task of 26.59% in ACC, 7.44% in SPE, 24.98% in REC, 28.97% in PRE, and 26.89% in F1. The enhancements observed in the five-class classification task are substantially greater compared to those in the binary classification task. The reduction in standard deviation in the prediction results also indicates that the inclusion of snRNA data significantly enhances the stability of scBIT's predictions. We conducted an evaluation to determine the impact of integrating different proportions of snRNA data on the predictive performance of scBIT. This assessment involved a systematic introduction of snRNA data at 20% intervals, ranging from 20% to 80%. The results (**Fig. 3(d)**) clearly demonstrated that the inclusion of incremental amounts of snRNA data significantly improved the accuracy of scBIT in binary classification task. Notably, the average accuracy increased from 0.9431 with 20% snRNA data to 0.9569 when 80% was included. Additionally, we evaluated the performance of scBIT across different demographic groups (gender and age). The results (**Fig. 3(b)** and **Fig. 3(c)**) indicate that scBIT achieves better predictive accuracy in the male group and the younger age group (ages 50-70), with an average ACC of 98.86 for the male group and 97.65 for the age 50-70 group.

In order to further assess the predictive capabilities of the scBIT model, we compared it against other existing methods specifically designed for fMRI data analysis. The comparison encompassed not only methodologies developed for AD utilizing the same dataset (ADNI) as employed in this work, but also approaches designed for the diagnosis of other neurological disorders. For the former, some methods show variable performance due to the integration of different fMRI representation strategies, therefore we selected the best results reported in the literature. For the latter, we have made adjustments to these methods to enable them to process the same fMRI dataset as ours. The comparison results presented in **Table 1** show that our scBIT method achieves the highest scores, significantly surpassing other competing models, including those designed for different diseases. Specifically, scBIT achieves an accuracy of 0.958, which is 11.66% higher than the average accuracy of methods targeting AD and 6.63% higher than those developed for other diseases. This superior performance emphasizes the effectiveness of scBIT in enhancing diagnostic accuracy for AD.

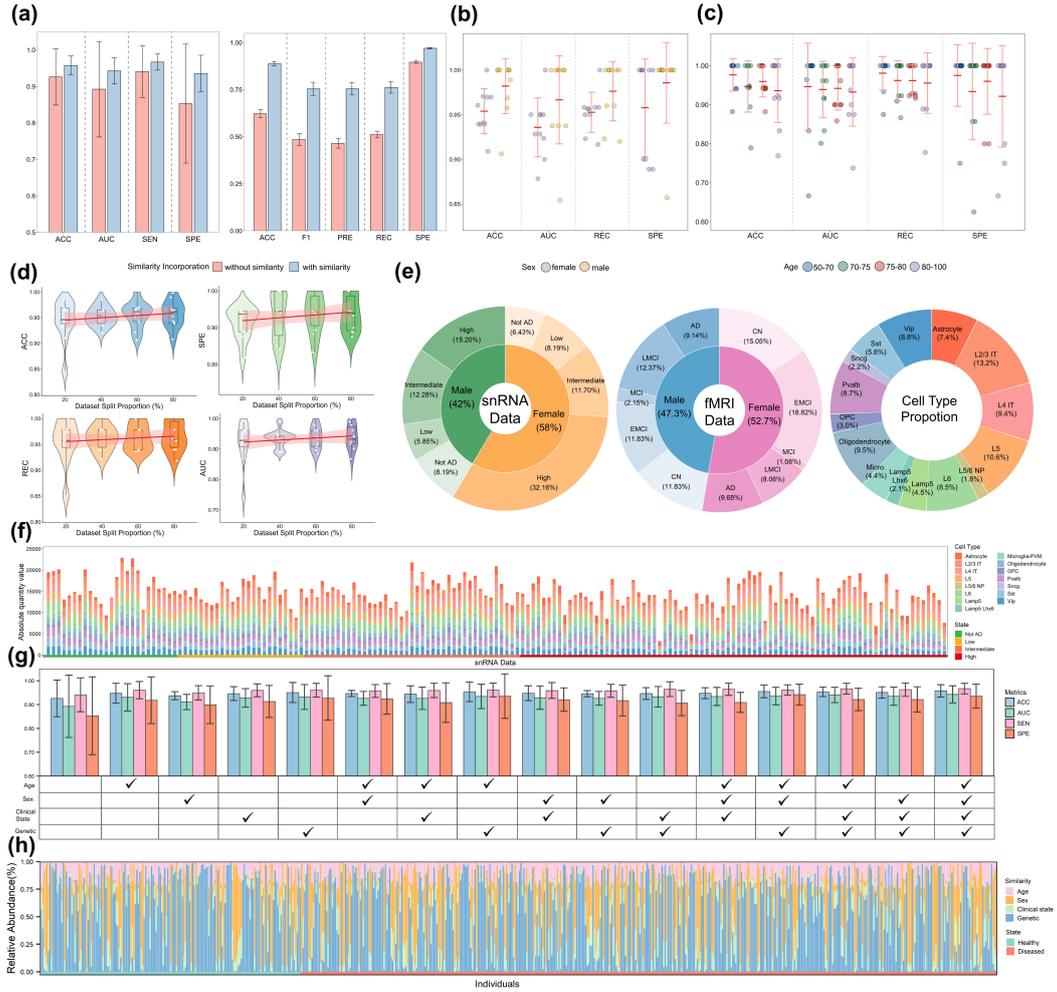

**Fig. 3.** Predictive Performance of the scBIT Model and Statistics of the snRNA-fMRI Cross-Modal Dataset. (a) Comparison of prediction performance with and without pre-trained embeddings in (left) binary disease diagnosis tasks and (right) classifying five severity levels of AD. (b) Disease diagnosis predictions under gender division. (c) Disease diagnosis predictions across different age groups. (d) Performance of scBIT with incremental introduction of snRNA data at 20% intervals. (e) Proportions of individuals based on disease status and gender in the snRNA and fMRI datasets, and the proportions of total cell numbers across different cell types in the snRNA dataset. (f) Cell counts of each cell type per individual in the snRNA dataset. (g) Performance comparison of scBIT in predicting AD diagnosis using different combinations of cross-modal individual similarities. (h) Proportions of automatically learned importance weights for four different similarities by MoE in scBIT using all similarities for each fMRI data point.

Table 1. Comparison of Prediction Performance Using 10-Fold Cross-Validation on the Binary Classification of AD Status in the ADNI Dataset.

| Method Type | Methods | ACC | SPE | SEN | AUC |
|---|---|---|---|---|---|
| Methods for AD | Chen's work [37] | 0.764 | 0.798 | 0.650 | 0.760 |
| | Bolla's work [38] | 0.850 | 0.710 | 0.930 | <u>0.930</u> |
| | Shi's work [39] | 0.929 | 0.867 | **1.000** | NA |
| | GP-LR [40] | 0.801 | <u>0.902</u> | 0.710 | NA |
| | Lama'work [41] | 0.863±0.033 | 0.901±0.049 | 0.823±0.065 | NA |
| Methods for other diseases | SGCOA_SCA [15] | 0.906±0.082 | 0.853±0.081 | 0.945±0.074 | 0.922±0.076 |
| | ST-GCN [42] | 0.857±0.059 | 0.806±0.116 | 0.881±0.109 | 0.892±0.026 |
| | LG-GNN [43] | <u>0.912±0.048</u> | 0.895±0.096 | 0.946±0.037 | 0.904±0.048 |
| Ours | scBIT | **0.958±0.026** | **0.936±0.050** | <u>0.967±0.022</u> | **0.942±0.036** |

The results are presented as mean±standard deviation. The best results are highlighted in bold, and the second-best results are underlined.

## 4.3 scBIT enables cross-modal contrastive learning for unpaired individuals

Due to the use of snRNA data from living individuals and fMRI data from a donor cohort for AD research, it is inherently unfeasible to collect cross-modal data from the same individual, resulting in unpaired sampling issues when constructing similarity measures for contrastive learning model frameworks. To address this challenge, scBIT harnesses the demographic and clinical information of data contributors by integrating four distinct attributes (i.e., gender, age, clinical condition, and genetics) to match datasets across two different modalities through the computation of similarity metrics. These metrics enable the training of two pre-trained regression models (for age and genetic similarities) and two pre-trained classification models (for clinical condition and gender), each incorporating an encoder framework. To assess the effectiveness of different individual similarities, we tested all combinations of embeddings from their respective encoders, employing a MoE model for integration. The results (**Fig. 3(g)**) show that genetic similarity achieved the highest average accuracy in single-encoder comparisons, reaching 95.05. Furthermore, the results reveal that incorporating more individual similarities leads to higher accuracy for scBIT. Average accuracies improve with the number of encoders used, ranging from 94.53 with a single encoder, 94.73 with two encoders, 95.19 with three encoders, to 95.75 with four encoders. **Fig. 3(h)** shows the proportion of weights for each encoder, as automatically determined by the MoE model, for each individual fMRI dataset. We quantitatively assessed the average weights of four types of individual similarity from all fMRI data, revealing proportions of 11.25% for age, 25.82% for sex, 11.11% for clinical state, and 51.82% for genetic similarity. The predominant weight for genetic similarity aligns with results from single-similarity metric tests, highlights the significant role of genetic factors in the integrated analysis of fMRI data.

## 4.4 scBIT provides interpretability of pathogenic gene networks across various gene pathway databases

scBIT uses a cell bagging strategy to build gene interaction networks from snRNA data, providing interpretability at the gene relationship level. To evaluate the impact of different gene network constructions on model performance, we introduced six gene pathway datasets. Specifically, we employed the gene lists from the gene pathway datasets as filtering criteria to ensure that the gene interaction networks extracted by scBIT focus on specific gene pathway datasets. The number of genes in these datasets varies from 779 to 12,979, with the statistical data presented in **Fig. 4(a)**. The results (**Fig. 4(b)**) demonstrate that scBIT's predictive performance is generally robust to the choice of pathway database, achieving the highest accuracy of 0.9615 with the Reactome dataset and the lowest accuracy of 0.9523 with the KEGG dataset. This suggests that scBIT can feasibly predict Alzheimer's diagnosis by integrating specific gene pathway datasets and provide interpretable gene relationships. Large gene pathway datasets, despite encompassing a broader array of genes, often lack targeted specificity; in contrast, smaller datasets, though limited in scope, exhibit greater specificity. The results indicate that there is no evident correlation between the number of genes in a dataset and the representational effectiveness of scBIT's snRNA feature extraction method, likely due to the trade-off between dataset size and specificity. We calculated the AUCell scores for all cells across 171 snRNA datasets using the smallest gene pathway dataset (HumanCyc), and grouped them into 15 major cell types. The distribution (**Fig. 4(c)**) revealed significant differences in gene pathway expression among each cell type, inspiring us to train the snRNA feature extraction network using cell types as labels. We further analyzed the average embedding representations for each snRNA dataset corresponding to 15 cell type subgraphs, and displayed their distribution in **Fig. 4(d)**. The results show that the embeddings for each cell type exhibited a pronounced clustering effect, indicating significant differentiation among different cell types. **Fig. 4(e)** illustrates the gene subgraphs acquired by scBIT for both the patient and healthy groups using the smallest pathway dataset. For each cell type, we averaged the data within the group and displayed the top 20 gene-gene interactions that distinguish between the patient and healthy individuals.

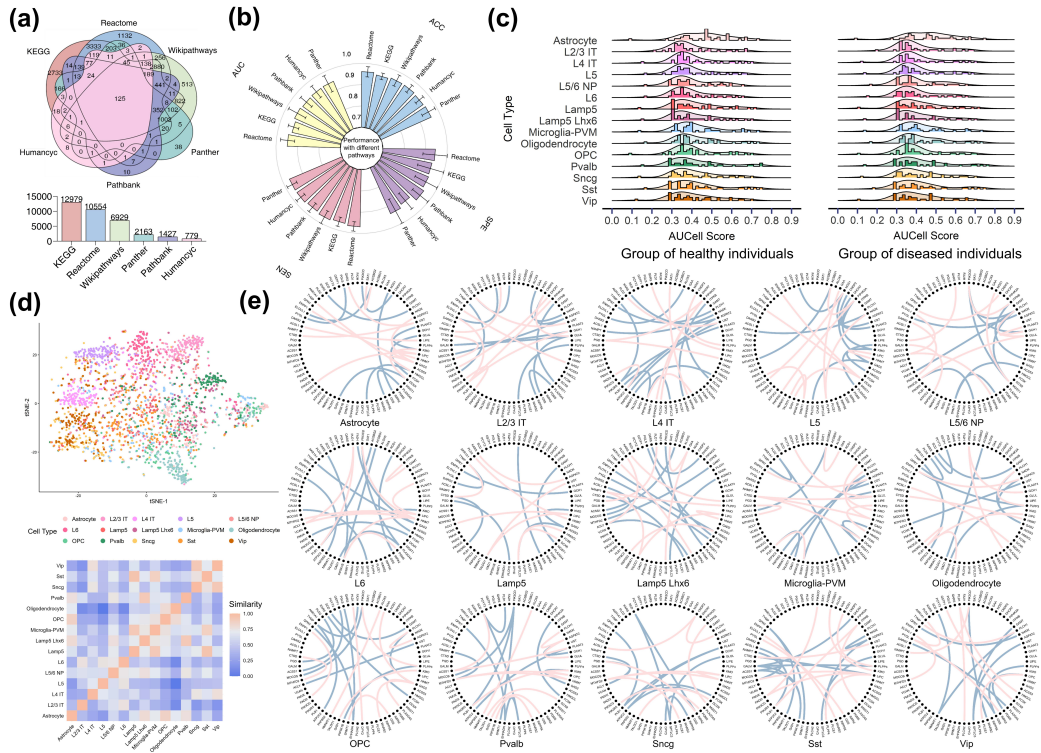

**Fig. 4.** Predictive performance of scBIT across different public gene pathway datasets. (a) Venn diagram of gene counts across six gene pathway datasets used. (b) Performance comparison of scBIT using different gene pathway databases. (c) Distribution of gene expression across different cell types in patient and healthy groups using the minimal gene pathway dataset for snRNA data. (d) Distribution of gene subgraph phenotypes for 15 cell types learned from all snRNA data and their respective cluster centroid distances. (e) Visualization of the most representative gene interaction subgraphs for different cell types as learned by scBIT.

## 4.5 scBIT offers interpretative insights into associations between brain regions, gene networks, and cell types

Utilizing the cross-attention mechanism [44], scBIT facilitates the integration of snRNA and fMRI modalities into a unified embedding space. This integrated embedding encapsulates the intricate interactions between cell types and brain regions. The attention weights derived from this model may identify connections between brain regions and genes, or between brain regions and cell types, as potential significant biomarkers for the mechanisms of AD. **Fig. 5(a)** visualizes the average attention weights for patients in the disease group, with the left half of the figure displaying connections between each cell type and the brain regions ranked in the top 20% of attention scores, and the right half showing connections from each brain region to the cell types also within the top 20% of attention scores, highlighting significant brain region-cell type associations. To evaluate the significant differences in brain region connections between the disease and healthy groups, we utilized the explainability technique Grad-CAM [45] to back-calculate and reconstruct the fMRI data inputs based on the diagnostic results from the trained scBIT model. Through this reconstructed fMRI data, we calculated the Pearson correlation coefficients (PCCs) between each ROI and averaged these coefficients across all samples, selecting the top 10 ROI connections with the highest PCCs. These connections are depicted in **Fig. 5(b)** and **Fig. 5(d)**, highlighting key functional connectivity patterns associated with the disease and providing important biomarkers for further research and diagnosis. We further evaluated the stability of subgraphs within the patient group. We identified the top five most stable subgraphs, derived from the minimal gene pathway dataset, characterized by the smallest variance in attention scores. These subgraphs, along with their corresponding brain regions exhibiting the highest attention scores, are shown in **Fig. 5(c)**.

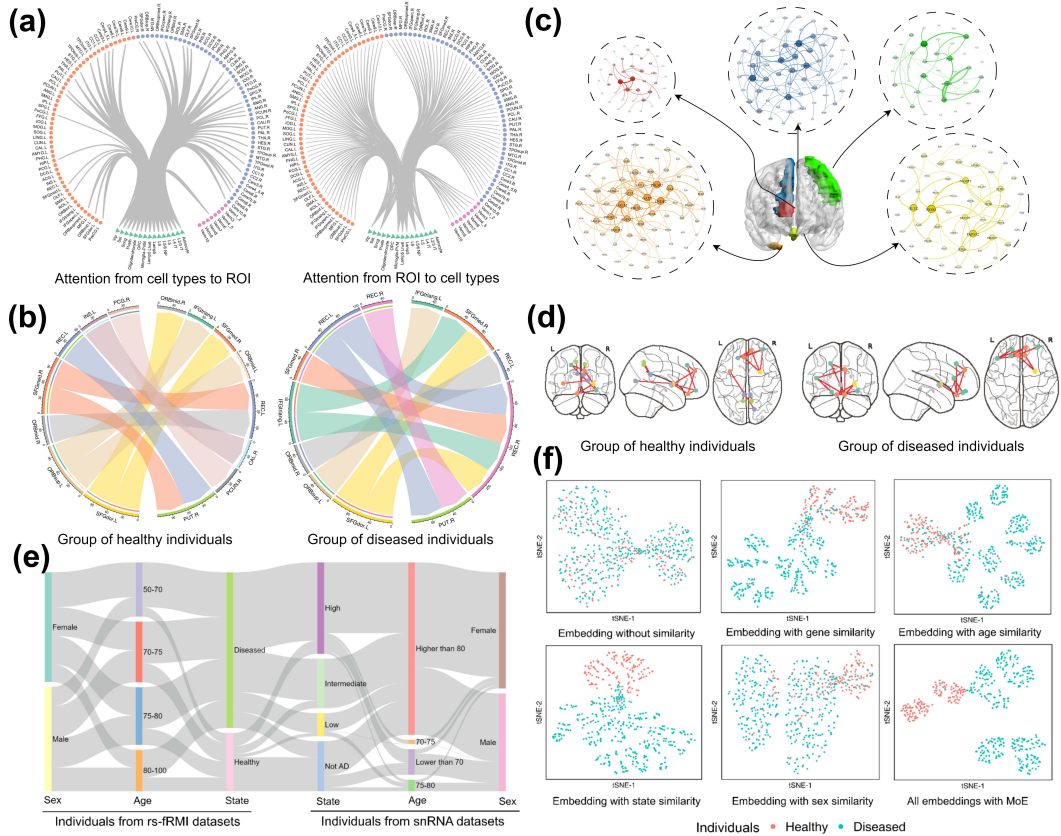

**Fig. 5.** scBIT provides interpretability regarding the correlations among cell type, brain region, and gene networks in AD. (a) Visualization of attention between cell types and brain regions in the patient group learned from scBIT, showing the top 20% of (left) brain regions with the highest attention for each cell type and (right) cell types with the highest attention for each brain region. (b) Circular graph visualization of the top 10 functional connectivities calculated using post-hoc interpretation techniques for healthy individuals (left) and patients (right). (c) Display of the top 5 most robust subgraphs learned from the minimal gene pathway dataset for the patient group and the linked brain regions exhibiting the highest attention. (d) 3D brain networks for the top 10 functional connectivities in healthy individuals (left) and patients (right). (e) Sankey diagram of cross-dataset patient correlations learned from scBIT. (f) Distribution of joint embeddings of each fMRI data combined with different pre-trained embeddings.

## 4.6 scBIT facilitates the identification of relevant cases across cross-modal datasets

In the scBIT model, snRNA data are initially transformed into an embedding set represented by gene interaction subgraphs, which serve as a dictionary of cases for retrieval. Given specific fMRI data, the scBIT employs a cross-attention mechanism to compute the individual-wise attention between the fMRI data and 171 snRNA data instances, thereby facilitating case matching across multimodal datasets. **Fig. 5(e)** displays the matching results for each fMRI sample in scBIT, highlighting the cases that have the highest individual-wise attention and are thus most closely related. It is noted that during the prediction phase, the model exclusively utilizes fMRI data, excluding personal demographic or genomic information, meaning that attention scores are derived solely from fMRI data. To further assess the influence of our pre-trained models, we plotted the embedding distributions for each fMRI dataset in **Fig. 5(f)**, including raw data embeddings, embeddings from four distinct pre-trained encoders, and embeddings from the MoE model. We conducted a quantitative assessment using the k-means clustering algorithm to calculate the Adjusted Rand Index (ARI) for these distributions, based on embeddings derived from single similarity measures of genomic data, age, clinical state, and sex. The resulting ARIs were 0.407, 0.372, 0.779, and 0.345, respectively. By integrating these four types of similarity through the MoE model, we achieved a significantly enhanced ARI of 0.9429.

## 5 Conclusions

The scBIT method represents a pioneering advancement in Alzheimer's disease (AD) research, effectively bridging the gap between neuroimaging and molecular genetics. By integrating functional magnetic resonance imaging (fMRI) with single-cell transcriptomics [46-48], scBIT not only enhances diagnostic accuracy but also deepens our understanding of AD's biological underpinnings. This integration facilitates the identification of novel biomarkers through the discovery of significant associations between specific brain regions and gene interaction subgraphs, thus offering a dual perspective on the pathophysiology of AD.

The multimodal approach employed by scBIT distinguishes it from existing methodologies by preserving cell-type heterogeneity and leveraging the unique strengths of both data types to unveil intricate patterns that are typically obscured when studied in isolation [49, 50]. The ability of scBIT to utilize attention mechanisms to highlight critical gene-brain region associations introduces potential novel biomarkers, positioning it as a transformative tool in the field.

Looking forward, the application of scBIT could revolutionize the re-analysis of existing datasets, enhancing the value of public single-cell transcriptomic and fMRI data. As we continue to refine this method, we anticipate it will encourage more comprehensive studies involving diverse patient demographics and disease stages, ultimately leading to advanced diagnostic tools and therapeutic strategies tailored to individual pathological profiles. This work underscores the immense potential of integrating diverse biomedical datasets to advance our understanding and treatment of complex diseases like AD.

# 6   Funding

This research was funded by the National Science Fund for Distinguished Young Scholars of China under grant number: 62325308; the Science and Technology Innovation 2030–New Generation Artificial Intelligence Major Project under grant number No. 2018AAA0100103; the National Natural Science Foundation of China under grant number 62002297, grant number 61722212, grant number 62072378, grant number 62273284, grant number 62472353, grant number 62302495, and grant number 62172338; the Neural Science Foundation of Shaanxi Province under grant number: 2022JQ-700; the Fundamental Research Funds for the Central Universities under grant number: D5000230199; the Guangdong Basic and Applied Basic Research Foundation under grant number 2024A1515011984; and the Fundamental Research Funds for the Central Universities under grant number G2023KY05102.

## 7 Declaration of competing interest

The authors declare that they have no conflicts of interest in this work.